\documentstyle[emlines2]{article}
\def\emline#1#2#3#4#5#6{%
       \put(#1,#2){\special{em:moveto}}%
       \put(#4,#5){\special{em:lineto}}}
\def\newpic#1{}

\def\hybrid{\topmargin 0pt      \oddsidemargin 0pt
        \headheight 0pt \headsep 0pt

       \textwidth 6.5in        
      \textheight 9in         
        \marginparwidth 0.0in
        \parskip 5pt plus 1pt   \jot = 1.5ex}
\catcode`\@=11
\def\marginnote#1{}

\newcount\hour
\newcount\minute
\newtoks\amorpm
\hour=\time\divide\hour by60
\minute=\time{\multiply\hour by60 \global\advance\minute by-\hour}
\edef\standardtime{{\ifnum\hour<12 \global\amorpm={am}%
        \else\global\amorpm={pm}\advance\hour by-12 \fi
        \ifnum\hour=0 \hour=12 \fi
        \number\hour:\ifnum\minute<10 0\fi\number\minute\the\amorpm}}
\edef\militarytime{\number\hour:\ifnum\minute<10 0\fi\number\minute}

\def\draftlabel#1{{\@bsphack\if@filesw {\let\thepage\relax
   \xdef\@gtempa{\write\@auxout{\string
      \newlabel{#1}{{\@currentlabel}{\thepage}}}}}\@gtempa
   \if@nobreak \ifvmode\nobreak\fi\fi\fi\@esphack}
        \gdef\@eqnlabel{#1}}
\def\@eqnlabel{}
\def\@vacuum{}

\def\draftmarginnote#1{\marginpar{\raggedright\scriptsize\tt#1}}

\def\draftlabel#1{{\@bsphack\if@filesw {\let\thepage\relax
   \xdef\@gtempa{\write\@auxout{\string
      \newlabel{#1}{{\@currentlabel}{\thepage}}}}}\@gtempa
   \if@nobreak \ifvmode\nobreak\fi\fi\fi\@esphack}
        \gdef\@eqnlabel{#1}}
\def\@eqnlabel{}
\def\@vacuum{}
\def\draftmarginnote#1{\marginpar{\raggedright\scriptsize\tt#1}}

\def\draft{\oddsidemargin -.5truein
        \def\@oddfoot{\sl preliminary draft \hfil
        \rm\thepage\hfil\sl\today\quad\militarytime}
        \let\@evenfoot\@oddfoot \overfullrule 3pt
        \let\label=\draftlabel
        \let\marginnote=\draftmarginnote
   \def\@eqnnum{(\theequation)\rlap{\kern\marginparsep\tt\@eqnlabel}%
\global\let\@eqnlabel\@vacuum}  }

\def\numberbysection{\@addtoreset{equation}{section}
        \def\theequation{\thesection.\arabic{equation}}}

\def\underline#1{\relax\ifmmode\@@underline#1\else
        $\@@underline{\hbox{#1}}$\relax\fi}

\def\titlepage{\@restonecolfalse\if@twocolumn\@restonecoltrue\onecolumn
     \else \newpage \fi \thispagestyle{empty}\c@page\z@
        \def\thefootnote{\fnsymbol{footnote}} }

\def\endtitlepage{\if@restonecol\twocolumn \else  \fi
        \def\thefootnote{\arabic{footnote}}
        \setcounter{footnote}{0}}  
\relax

\numberbysection
\hybrid

\def\beq{\begin{equation}}
\def\eeq{\end{equation}}
\def\p{\partial}
\def\G{\Gamma}

\newtheorem{th}{Theorem}[section]

\newtheorem{lem}{Lemma}[section]

\begin{document}

\begin{titlepage}

\title{Elliptic solutions to difference non-linear equations
and related many-body problems}

\author{I. Krichever \thanks{Department of Mathematics of Columbia
University and Landau
Institute for Theoretical Physics
Kosygina str. 2, 117940 Moscow, Russia}
\and P. Wiegmann \thanks{James Franck Institute and
and Enrico Fermi Institute of the University of Chicago, 5640 S.Ellis
Avenue, Chicago, IL 60637, USA and
Landau Institute for Theoretical Physics}
\and A. Zabrodin
\thanks{Joint Institute of Chemical Physics, Kosygina str. 4, 117334,
Moscow, Russia and ITEP, 117259, Moscow, Russia}}

\maketitle

\begin{abstract}

We study algebro-geometric (finite-gap) and elliptic solutions 
 of fully discretized  
KP or 2D Toda equations. In bilinear form they are  
 Hirota's difference equation for $\tau$-functions. 
 Starting from a given algebraic curve,
 we express the $\tau$-function
  and the Baker-Akhiezer function in terms of the Riemann theta function.
 We show that the elliptic solutions,
  when the $\tau$-function is an elliptic polynomial,
form a subclass of the general algebro-geometric solutions.
We construct the
algebraic curves of the elliptic solutions.
The evolution of zeros of the elliptic solutions is governed by 
the discrete time generalization of the
Ruijsenaars-Schneider many body system. The zeros obey equations which
have the 
form of nested
Bethe-Ansatz equations, known from integrable quantum field theories. 
We discuss the Lax representation 
and the action-angle-type
variables for the many body system. We also discuss elliptic solutions to
discrete analogues of KdV, sine-Gordon
and 1D Toda equations and describe the loci of the zeros.

\end{abstract}

\vfill

\end{titlepage}

\section{Introduction}

Among a vast class of solutions  to classical non-linear integrable
equations elliptic solutions play a special role. First, these
 solutions occupy
a distinguished place among all algebro-geometric
(also called finite-gap) solutions, i.e. solutions 
constructed out of a given algebraic curve. The general formulas in terms
of Riemann theta-functions become much more effective -- in this case the
Riemann theta-function splits  into a product of
Weierstrass $\sigma$-functions associated to an elliptic curve.
Second, there exists a remarkable connection between the motion
of poles (zeros) of the elliptic solutions and certain integrable many
body
systems. 

The pole dynamics of elliptic solutions to the Korteweg - de Vries
(KdV) equation and the Calogero-Moser system of particles were
linked together in the paper \cite{AKM} (see also \cite{chood}).
It has been shown in \cite{kr1},\cite{kr2} that this relation
becomes an isomorphism if one considers elliptic solutions
of the Kadomtsev-Petviashvili (KP) equation. More recently, these
results were generalized to elliptic solutions of the matrix KP
and the matrix 2D Toda lattice equations (see \cite{bab}
and \cite{kz}, respectively). The dynamics of their poles 
obeys the spin generalization of the
Ruijsenaars-Schneider (RS) model \cite{RS}.

Let us recall some elements of the elliptic solutions  
for the standard example of the KP
equation $3u_{yy} =(4u_{t}+6uu_{x}-u_{xxx})_{x}$ for a
function $u=u(x,y,t)$. An elliptic solution
 in the variable $x$ is given by
\beq
u(x,y,t)=\mbox{const}\; +2\sum _{i=1}^{N}\wp (x-x_{i}(y,t)),
\label{I0}
\eeq
where $\wp (x)$ is the Weierstrass $\wp$-function.
The self-consistency of this Ansatz is a
manifestation of integrability.
It has been shown in \cite{kr1},\cite{kr2} that the 
dynamics of poles as functions of $y$ obeys the Calogero-Moser
many body system with the 
Hamiltonian 
\beq
H=\frac{1}{2}\sum _{i=1}^{N}p_{i}^{2}-2\sum _{i\neq j}
\wp (x_{i}-x_{j}).
\label{I1}
\eeq
This system in its turn is known to be integrable. 
There is an involutive set of 
conserved quantities
$H^{(j)}$ -- the Hamiltonian (\ref{I1}) and the total momentum are 
$H^{(2)}$ and $H^{(1)}$. The equations of motion are
\beq
\p _{y}^{2}x_i =4\sum _{j=1, \neq i}^{N} \wp ' (x_i -x_j ).
\label{I2}
\eeq
The $t$-dynamics is described by $H^{(3)}$. 
 
The reduction to the KdV equation restricts the particles  to
the {\it locus} ${\bf L}_{N}$ in the phase space:
\beq
{\bf L}_{N}=\left \{\left. \phantom{\frac{a}{b}} \!\!
(p_i , x_i )\;\right |\;p_i =0, \; \sum _{i\neq j}\wp ' (x_i -x_j )=0
\; \right \}
\label{locus}
\eeq
(here $p_i = \p _{y}x_i$).
In spite of interesting developments, an analysis
of the locus structure is far from to be completed.
 
In this paper we extend these 
results to the fully discretized version
of the KP equation or 2D Toda lattice. Being fully discretized they become 
the same equation. In bilinear form they are
known as Hirota's
bilinear difference equation (HBDE) \cite{Hirota} (see \cite{Zabr} for a
review). 
This is a bilinear equation for a function
$\tau (l,m,n)$ (called $\tau$-function) of three variables:
\begin{eqnarray}
&&\lambda \tau (l+1 , m,n)\tau (l,m+1, n+1)
+\mu \tau (l , m+1,n)\tau (l+1,m, n+1)\nonumber \\
&+&\nu \tau (l , m,n+1)\tau (l+1,m+1, n)=0,
\label{I3}
\end{eqnarray}
where $\lambda , \,\mu ,\,\nu$ are complex parameters and the  
three  variables are not necessarily integer. In what follows we 
call them {\it discrete times} stressing the difference with continuous
KP-flows. Let us introduce a lattice spacing
$\eta$ for one of the variables, say, $l$ and denote $x\equiv \eta l$. 
 By {\it elliptic
solutions} (in the variable $x$) to this equation we mean the following 
Ansatz for the
$\tau$-function:
\beq
\tau (l,m,n)\equiv \tau ^{m,n}(x)=
\prod _{j=1}^{N}\sigma (x-x_{j}^{m,n}),
\label{I4}
\eeq
where $\sigma (x)$ is the Weierstrass $\sigma$-function. We refer to
the r.h.s. of (\ref{I4}) as {\it elliptic polynomials}
in $x$. For brevity, we call solutions of this type elliptic
though the $\tau$-function itself is not double-periodic. However,
suitable ratios of these $\tau$-functions, for instance,
\beq \label{I5}
A^{m,n}(x)=\frac{\tau ^{m,n}(x)\tau ^{m+1, n}(x+\eta )}
{\tau ^{m+1,n}(x)\tau ^{m, n}(x+\eta )}
\eeq
are already elliptic functions. 

In this paper we derive equations of motion 
for poles of
$A^{m,n}(x)$ or zeros of $\tau ^{m,n}(x)$ for discrete times $m,n$ 
and thus obtain a fully 
discretized Calogero-Moser many body problem. This appears to be the
discrete 
time version of the 
Ruijsenaars-Schneider (RS) model proposed in the seminal paper \cite{NRK}.
Remarkably, the discrete equations of motion
have the form of Bethe equations of the  hierarchical
(nested) Bethe Ansatz. The discrete time runs over "levels"
of the nested Bethe Ansatz.

We also consider  stationary reductions of HBDE. In this case the initial
 configuration of
poles (zeros) is not arbitrary but constrained to a 
stable locus as in the continuous case
(\ref{locus}). For the most important examples we give equations defining
the loci. 

A renewed interest in soliton difference equations, 
and especially in their
elliptic solutions is caused 
 by the  revealing {\it classical} integrable structures
present in  integrable models of {\it quantum} field theory. 
It turns out that Hirota's equation (\ref{I3})
 is the universal fusion rule for a family of  
quantum transfer matrices. Their eigenvalues 
(as functions of spectral parameters)
obey a set of functional equations
\cite{Kuniba} which can be recast into the bilinear
 Hirota equation
 \cite{KLWZ} (see also \cite{Z} and \cite{Wiegmann} for less technical
reviews).
Furthermore, it turned out that most of the ingredients of
the Bethe Ansatz and the quantum inverse scattering method 
 are hidden in the  elliptic 
solutions of the entirely classical discrete time soliton equations
\cite{KLWZ}. In particular, the discrete dynamics of poles of $A^{m,n}(x)$
or zeros of (\ref{I4}) has the form of Bethe Ansatz equations, 
where the discrete time runs over
"nested" levels.

The theory of elliptic solutions has direct
applications to the algebraic Bethe Ansatz and to Baxter's 
$T$-$Q$-relation, which we plan to
discuss elsewhere.

Here we attempt to develop a  systematic approach to the 
elliptic solutions of the integrable difference 
 equations. The basic concept of the approach is 
the Baker-Akhiezer functions on algebraic curves.
We prove that all solutions to HBDE of the form
(\ref{I4}) are
of the algebro-geometric type and present them in terms of 
Riemann theta functions.

The plan of the paper is as follows.

In  Sect.\,2 we describe  general algebro-geometric (finite-gap)
solutions to HBDE.
We start from  the Baker-Akhiezer function
constructed from a complex algebraic curve
of genus $g$ with marked points. 
This function satisfies an overcomplete set of linear
difference equations.
Their consistency is equivalent to Hirota's equation. 
In this way, one obtains a
$(4g+1)$-parametric family of quasiperiodic solutions to HBDE
in terms of the Riemann theta-functions.
Solitonic degenerations of these solutions are  discussed in Sect.\,2.4.

Sect.\,3 is devoted to elliptic solutions. They are shown to
be a particular subclass of the algebro-geometric 
family of solutions of Sect.\,2.
We derive equations
of motion for zeros of the $\tau$-function 
(the Bethe Ansatz equations) and their 
Lax representation. 
We discuss variables of the action-angle type and 
write down equations for the stable
loci for the most important reductions of Hirota's equation. 

\section{Algebro-geometric solutions to Hirota's equation}

In this section we construct algebro-geometric solutions
of Hirota's equation out of a given algebraic curve.
The general method of constructing such solutions of 
 integrable equations 
is standard. As soon as the
bilinear equation can be represented as a compatibility condition
for an overdetermined system of linear problems, the first step
is to pass to common solutions
$\Psi$ to the {\it linear} problems. Given a linear
multi-dimensional difference operator with quasiperiodic
coefficients, one associates with it a
{\it spectral curve}
defined by the generalized dispersion relations for
quasimomenta of Bloch
eigenfunctions of the linear operator. The
Bloch solutions $\Psi$ are parametrized by points of this
curve. Solutions to the initial non-linear equation are encoded
in the analytical properties of $\Psi$ as a function
on the curve.
Spectral curves of general linear operators with
quasiperiodic coefficients are transcendental and,
therefore, intractable. However, soliton theory
mostly deals with the {\it inverse} problem:
to characterize specific operators whose spectral
curves are {\it algebraic curves of finite genus}.
Such operators are called algebro-geometric or finite-gap.
Their coefficients yield solutions to HBDE, that we are
going to study.

Finite-gap multi-dimensional linear
difference operators were constructed
in the paper \cite{kr10} by one of the authors. We present the
corresponding construction in a form adequate for
our purposes.

\subsection{The Baker-Akhiezer function}

As usual, we begin with the axiomatization of
analytical properties of Bloch solutions $\Psi$.
The Baker-Akhiezer function is an abstract version of
the Bloch function.
Since we solve the inverse problem, the primary objects
are $\Psi$-functions rather than linear operators.

Let $\G$ be a smooth algebraic curve of genus $g$.
We fix the following data related to the curve:
\begin{itemize}
\item[-]{A finite set of marked points (punctures)
$P_{\alpha}\in \G$, $\alpha = 0,1,\ldots , M$;}
\item[-]{Local parameters $w_{\alpha}$ in neighbourhoods of 
$P_{\alpha}$:
$w_{\alpha}(P_{\alpha})=0$;}
\item[-]{A set of cuts $C_{\alpha \beta}$
between the points $P_{\alpha}, P_{\beta}$
for some pairs $\alpha, \beta$
(it is implied that
different cuts do not have common points other than their
endpoints at the punctures);}
\item[-]{A set $D$ of $g$ (distinct) points 
$\gamma _{1}, \ldots , \gamma
_{g} \in \G$.}
\end{itemize}

\noindent
Further, we introduce the following complex parameters $l_{\alpha\beta}$
 (times or flows):
\begin{itemize}
\item[-]{To each cut
$C_{\alpha \beta}$ is associated a complex number
$l_{\alpha \beta}$ (it is convenient to assume that
$l_{\beta \alpha}=-
l_{\alpha \beta}$).}
\end{itemize}

Consider a linear space ${\cal F}(l;D)$ of
functions $\Psi(l;P), \ P\in \G$, such that:
\begin{itemize}
\item[$1.$] {The function
$\Psi(l;P)$ as a function of the variable $P\in \Gamma$
is meromorphic outside the cuts and has at most simple poles at the points
$\gamma_s$;}
\item[$2.$] {The boundary
values $\Psi^{\pm,(\alpha \beta )}$ of this function
at opposite sides of the cut $C_{\alpha \beta}$ satisfy the relation
\beq
\Psi^{+,(\alpha \beta )}(l;P)=\Psi^{-,(\alpha \beta )}(l;P)
e^{2\pi il_{\alpha \beta}};
\label{3.1}
\eeq   }
\item[$3.$] {In a
neighbourhood of the point $P_{\alpha}$ it has the form
\beq
\Psi(l;P)=w_{\alpha}^{-L_{\alpha}}
\left(\xi_{0}^{(\alpha )}(l)+\sum_{s=1}^{\infty} \xi_{s}^{(\alpha )}(l)
w_{\alpha}^s \right), \;\;\;\;\;\;\;\;
L_{\alpha }=\sum _{\beta}l_{\alpha \beta}.
\label{3.2}
\eeq }
\end{itemize}
Note that if $l_{\alpha \beta}$ are integers,
then $\Psi$ is a meromorphic function
having simple poles at $\gamma_s$ and having
zeros or poles of
orders $|L_{\alpha }|$
at the points
$P_{\alpha }$.

Any function $\Psi$ obeying conditions $1$\,-\,$3$ is called a
{\it Baker-Akhiezer function}. In our approach, these functions
are central objects of the theory.
In some cases (especially in the matrix
generalizations of the theory) the notion of the {\it dual
Baker-Akhiezer function} $\Psi ^{\dagger}$ is also important
(see e.g.\,\cite{bab}, \cite{kz}). We omit its definition
because it can be easily restored using
\cite{bab}, \cite{kz}.

Let us prepare some notation. Fix a canonical basis of cycles
$a_i , b_i$ on $\G$ and denote the canonically normalized holomorphic
differentials by $d\omega _{i}$, $i=1,2, \ldots ,g$.
We have
$$
\oint _{a_i}d\omega _{j}=\delta _{ij}\,,
\;\;\;\;\;\;\;\;\;
\oint _{b_i}d\omega _{j}=B_{ij}\,,
$$
where $B$ is the period matrix. Given a period matrix $B$,
the $g$-dimensional Riemann
theta-function is defined by
$$
\Theta (\vec X )=
\Theta (\vec X | B)=
\sum _{\vec n \in {\bf Z}^{g}}
\exp \left [\pi i (B\vec n , \vec n )
+2\pi i (\vec n , \vec X) \right ].
$$
Here $\vec X =(X_1 , \ldots , X_g )$ is a $g$-component vector.

For each pair of points
$P_{\alpha}, P_{\beta } \in \G$, let  $d\Omega ^{(\alpha \beta )}$ be
the unique differential of the third kind 
 holomorphic
on $\G$ but  having simple poles at the  points   $P_{\alpha}$,
$P_{\beta}$
 with  residues 
 $-1$ and $1$ and  zero $a$-periods.

To write an  explicit form of Baker-Akhiezer functions, let us choose
one of the marked points, say $P_0$, and by $\vec A(P)=(A_1(P),\ldots
\,, A_g(P))$, 
$$
A_i(P)=\int_{P_0}^P d\omega_i,
$$
denote the Abel map. The Baker-Akhiezer function is given by
the following theorem.
\begin{th}
If the points $\gamma _{1}, \ldots , \gamma _{g}$ are
in a general position (i.e. $D$ is a non-special divisor), then
${\cal F}(l;D)$ is a one-dimensional space generated by the
function $\Psi(l;P)\in {\cal F}(l;D)$
\beq
\Psi(l;P)\  =
\ {\Theta(\vec A(P)+\vec X(l) +\vec Z|B) \Theta (\vec Z|B) \over
\Theta (\vec A(P)+\vec Z|B) \Theta (\vec X(l)+\vec Z|B) }
\exp \left (
\sum _{(\alpha \beta )}l_{\alpha \beta}
\int ^{P}_{Q_{0}}d\Omega ^{(\alpha \beta )}
\right ).
\label{3.3}
\eeq
Here
$Q_{0}\in \G$  is an arbitrary point 
in the vicinity of $P_{0}$, $Q_{0}\neq P_{0}$,
belonging to the integration path $P_{0}\rightarrow P$ in
the Abel map.
 Further,
\beq
\vec Z=-\vec K -\sum_{i=1}^g \vec A(\gamma_i),\;\;\;\;\;\;
\vec X(l)=\sum _{(\alpha \beta )}
\vec U^{(\alpha \beta )}l_{\alpha \beta},
\label{3.6}
\eeq
where  $\vec K$ is the vector of Riemann's constants and 
the components of the vectors
$\vec U^{(\alpha \beta )}$
are
\beq
(\vec U^{(\alpha \beta )})_{j}=
\frac{1}{2\pi i}\oint _{b_j}d\Omega ^{(\alpha \beta)}=
A_{j}(P_{\beta})-A_{j}(P_{\alpha}).
\label{3.7}
\eeq
\end{th}

\noindent
The proof of theorems of this kind as well as the explicit  formula for
$\Psi$ in terms of Riemann theta-functions are standard in 
finite-gap theory (see e.g. \cite{kr}).
The last equality in (\ref{3.7})
follows from Riemann's relations.

\noindent
{\bf Remark 2.1}\,\,Although the explicit formula (\ref{3.3}) requires
a fixed basis of cycles, the Baker-Akhiezer function is modular invariant.

\noindent
{\bf Remark 2.2}\,\,Since abelian integrals
in (\ref{3.3})
have logarithmic singularities at the punctures,
one can define a single valued
branch of $\Psi$ only after cutting the curve along
$C_{\alpha \beta}$.

\noindent
{\bf Remark 2.3}\,\,The choice of the initial point of
the Abel map  is in fact not essential. It can be 
chosen not to be one of the punctures.
 In particular, it may be $Q_0$, which
 slightly simplifies the theorem. However,
our choice simplifies the linear  equations for $\Psi$ below.
Moreover, below we assume that the integration paths in the
Abel maps $\vec A(P_{\alpha})$ go along the cuts $C_{0\alpha}$.

\noindent
{\bf Remark 2.4}\,\,The theorem implies that any function from
$ {\cal F}(l;D)$ has the form
$r(l)\Psi(l;P)$, where $r(l)$ is an arbitrary function of $l$  but does
not 
depend on $P$.
It is convenient to choose it
 such  that at the point $P_0$ the first
regular term    $\xi _{0}^{(0)}(l)$ in (\ref{3.2})
equals 1.

Coefficients $\xi_s^{(\alpha)}$ of the asymptotical behaviour of the
$\Psi$
(\ref{3.2}) can be expressed through the $\tau$-function
\beq\label{F}
\tau(l)\equiv\tau(\{l_{\alpha\beta}\})=\Theta(\vec X(l)+\vec Z).
\eeq
In particular,
\beq\label{xi}
\frac{\xi_0^{(\alpha)}(l)}{\xi_0^{(\beta)}(l)}=\chi_{\alpha\beta}
\frac{\tau(l_{0\alpha}+1,l_{0\beta})}{\tau(l_{0\alpha},l_{0\beta}+1)},
\;\;\alpha,\beta\neq 0,
\eeq
where $\chi_{\alpha\beta}$ are $l$-independent constants.
Here and thereafter we skip unshifted arguments.

\noindent{\bf Remark 2.5}\,\,
If the graph of cuts includes a closed cycle, then a shift of variables
 $l_{\alpha \beta}\rightarrow l_{\alpha\beta}+1$ does not change the
 $\tau$-function but multiplies the $\Psi$-function by a cycle dependent
constant.
For instance, if the cycle consists of three links
$C_{\alpha\beta},\,C_{\beta\gamma},\,
C_{\gamma\alpha}$, then
\beq
\begin{array}{l}
\tau(l_{\alpha\beta}+1,l_{\beta\gamma}+1,l_{\gamma\alpha}+1)=
\tau(l_{\alpha\beta},l_{\beta\gamma},l_{\gamma\alpha}),
\\ \\
\Psi(l_{\alpha\beta}+1,l_{\beta\gamma}+1,l_{\gamma\alpha}+1;P)=
\mbox{const}\,\Psi(l_{\alpha\beta},l_{\beta\gamma},l_{\gamma\alpha};P).
\end{array}
\label{100}
\eeq
This follows from (\ref{3.1}), (\ref{3.3}).

In the sequel, we do not need the above construction in its full
generality. For our purposes it is enough to consider the case
of four punctures $P_{0}, \ldots , P_{3}$ and a general graph of cuts as
is in the 
figure. Cuts connect each pair of points. Any three links (not forming
 a  cycle) give rise to a bilinear equation of the Hirota type. They have
different
 forms, but are in fact equivalent due to (\ref{100}).
For further convenience we specify
\beq\label{I}
l_{01}=l_1,\;\;\;\;l_{02}=l_2,\;\;\;\;l_{03}=l_3,\;\;\;\;l_{12}=\bar l_3.
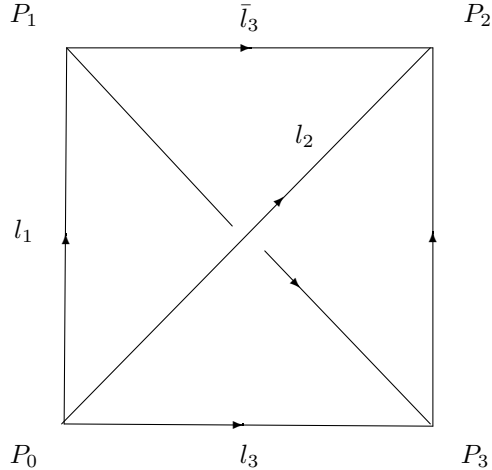
\eeq

\vspace{0.3cm}

\begin{center}
\special{em:linewidth 0.4pt}
\unitlength 1.00mm
\linethickness{0.4pt}
\begin{picture}(65.67,64.33)
\emline{10.67}{10.00}{1}{59.67}{9.67}{2}
\emline{59.67}{9.67}{3}{59.67}{60.00}{4}
\emline{59.67}{60.00}{5}{11.00}{60.00}{6}
\emline{11.00}{60.00}{7}{10.67}{10.00}{8}
\emline{10.33}{10.00}{9}{59.33}{60.00}{10}
\emline{11.00}{60.00}{11}{33.00}{36.33}{12}
\emline{59.33}{10.00}{13}{37.33}{33.00}{14}
\put(35.33,60.00){\vector(1,0){0.2}}
\emline{35.00}{60.00}{15}{35.33}{60.00}{16}
\put(59.67,35.33){\vector(0,1){0.2}}
\emline{59.67}{35.00}{17}{59.67}{35.33}{18}
\put(39.67,40.00){\vector(1,1){0.2}}
\emline{39.00}{39.33}{19}{39.67}{40.00}{20}
\put(42.67,48.33){\makebox(0,0)[cc]{$l_2$}}
\put(47.67,26.33){\makebox(0,0)[cc]{}}
\put(35.33,5.67){\makebox(0,0)[cc]{$l_3$}}
\put(35.33,64.00){\makebox(0,0)[cc]{$\bar l_3$}}
\put(5.33,35.67){\makebox(0,0)[cc]{$l_1$}}
\put(65.33,35.67){\makebox(0,0)[cc]{}}
\put(5.33,5.67){\makebox(0,0)[cc]{$P_0$}}
\put(65.33,5.67){\makebox(0,0)[cc]{$P_3$}}
\put(5.33,64.33){\makebox(0,0)[cc]{$P_1$}}
\put(65.67,64.33){\makebox(0,0)[cc]{$P_2$}}
\put(10.86,35.09){\vector(0,1){0.2}}
\emline{10.86}{35.02}{21}{10.86}{35.09}{22}
\put(34.32,9.86){\vector(1,0){0.2}}
\emline{34.10}{9.86}{23}{34.32}{9.86}{24}
\put(41.81,28.31){\vector(1,-1){0.2}}
\emline{41.63}{28.50}{25}{41.81}{28.31}{26}
\end{picture}
\end{center}

\vspace{0,2cm}

\noindent
The general case of more punctures yields higher Hirota equations (i.e.
 the discretized KP or 2D Toda lattice hierarchies).
 
\subsection{Difference equations for the Baker-Akhiezer function}

The Baker-Akhiezer function $\Psi (l;P)$
obeys certain linear difference
equations with respect to the variables $l_{\alpha \beta}$.
Coefficients of the equations are fixed by the analytical
properties of $\Psi (l;P)$ as a function of $P \in \G$. We
restrict ourselves to  the case of four punctures
and use the notation introduced
at the end of the previous subsection.
The general case of more punctures
can be treated in a similar way.

\begin{th}
Let $\Psi (l;P)$ be the Baker-Akhiezer function  normalized so that
$\xi _{0}^{(0)}=1$.
Then it satisfies the following linear difference equations:
\beq
\Psi (l_{\alpha}+1,l_{\beta};P)
-\Psi (l_{\alpha},l_{\beta}+1;P)+
A_{\alpha \beta}(l_{\alpha}, l_{\beta})
\Psi (l_{\alpha},l_{\beta};P)=0
\label{3.10}
\eeq
with
\beq
\label{3.11}
A_{\alpha \beta}(l_{\alpha}, l_{\beta})=\frac{
\xi _{0}^{(\alpha )}(l_{\alpha}, l_{\beta}+1)}
{\xi _{0}^{(\alpha )}(l_{\alpha}, l_{\beta})}
\eeq
for any $\alpha , \beta =1,2,3$, $\alpha \neq \beta$.
\end{th}
The proof is standard in finite gap theory. Denote the l.h.s.
of eq.\,(\ref{3.10}) by $\tilde \Psi$. This function has the same
analytical properties as $\Psi$. At the same time the leading
term at the point $P_1$ is zero: $\tilde \xi _{0}^{(1)}=0$ for
any $l_{\alpha}, l_{\beta}$. From the uniqueness of the Baker-Akhiezer
function it follows that $\tilde \Psi =0$. Eqs.\,(\ref{3.12}),
(\ref{3.12a}) are
proved in the same way.

\noindent
{\bf Remark 2.6}\,\,The dual Baker-Akhiezer function
$\Psi ^{\dagger}$
obeys difference equations obtained from eqs.\,(\ref{3.10}),
(\ref{3.12}) by conjugating the difference operators in the
right hand sides.

The coefficient functions of eq. (\ref{3.10}) are given by the 
leading nonsingular term $\xi_0^{(\alpha )}$ of the Baker-Akhiezer
function
at the punctures. They can be found from 
eq.\,(\ref{3.3}) and are expressed through the $\tau$-function (\ref{F}):
\beq
A_{\alpha \beta}(l_{\alpha}, l_{\beta})=
-\lambda _{\alpha \beta}
\frac{\tau (l_{\alpha},l_{\beta})\tau (l_{\alpha}+1,l_{\beta}+1)}
{\tau (l_{\alpha},l_{\beta}+1)\tau (l_{\alpha}+1,l_{\beta})}.
\label{3.14}
\eeq
The constants
$\lambda _{\alpha \beta}$
are expressed through the constant
terms $r_{\gamma}^{(\alpha \beta )}$ in expansion of the abelian
integrals
\beq
\left. \int _{Q_0}^{P} d\Omega ^{(\alpha \beta )}
\right | _{P\rightarrow P_{\gamma}}=
(\delta _{\gamma \beta }
-\delta _{\gamma \alpha })\log w_{\gamma}
+r_{\gamma}^{(\alpha \beta )}
+O(w_{\gamma}),
\label{3.16}
\eeq
as follows:
\beq
\lambda _{\alpha \beta}=-\exp \left (
r_{\alpha}^{(0 \beta )}-
r_{0}^{(0 \beta )}\right ).
\label{3.17}
\eeq
It can be shown that $\lambda _{\beta \alpha}=-
\lambda _{\alpha \beta}$ for $\alpha \neq \beta$ and
\beq
\frac{\lambda _{\alpha \beta}}{\lambda _{\beta \gamma}}=
-\exp \left (\int _{P_{\gamma}}^{P_{\alpha}}d\Omega ^{(0\beta)}\right )
\label{3.17a}
\eeq
for any cyclic permutation of $\{\alpha \beta \gamma \}
=\{123\}$. The integration path goes from $P_{\gamma}$ to the
neighbourhood of $P_0$ along the cut $C_{0\gamma}$ (in the
opposite direction), passes through the point $Q_0$ and then goes
along the cut $C_{0\alpha}$.

Eqs.\,(\ref{3.10}),\,(\ref{3.11}) can be viewed as  linear problems for
the
discretized KP equation. Choosing another triplet of variables, 
say $l_1,l_3,\bar l_3$ and 
using eq.\,(\ref{100}), i.e.
\beq
\tau (l_1 +1, l_2 , l_3 ;\bar l_3 +1)=
\tau (l_1 , l_2 +1, l_3 ;\bar l_3 ),
\label{dep}
\eeq
\beq
\tau (l_1 , l_2 , l_3; \bar l_{3})=
\Theta \left ( \sum _{\alpha =1}^{3} \vec A(P_{\alpha })l_{\alpha}+
\big (\vec A(P_{2})-\vec A(P_{1})\big )\bar l_{3}+\vec Z\right ),
\label{tau}
\eeq
one may rewrite eqs.\,(\ref{3.10}), (\ref{3.11}) in a form suitable
 for discretization of the 2D Toda lattice. In this case eq. (\ref{3.10}) 
for $\alpha,\beta=1,3$ remains the same. Another linear equation, 
\beq
\Psi (l_{1},\bar l_{3};P)
-\Psi (l_{1},\bar l_{3}+1;P)+B(l_{1}, \bar l_{3})
\Psi (l_{1}-1,\bar l_{3};P)=0\,,
\label{3.12}
\eeq
with
\beq
\label{3.13}
B(l_{1}, \bar l_{3})=\frac{
\xi _{0}^{(1)}(l_{1}, \bar l_{3}+1)}
{\xi _{0}^{(1)}(l_{1}-1, \bar l_{3})}=
-\lambda _{12}
\frac{\tau (l_{1}+1;\bar l_{3}+1)\tau (l_{1}-1;\bar l_{3})}
{\tau (l_{1};\bar l_{3}+1)\tau (l_{1};\bar l_{3})}
\label{3.15}
\eeq
follows from (\ref{3.10}) for $\alpha,\beta=1,2$ as a 
result of the change of variables.
The third equation,
\beq
\Psi (l_3 +1 , \bar l_{3}+1;P)
-\tilde A (l_3 , \bar l_{3})
\Psi (l_3 , \bar l_{3}+1;P)
=\Psi (l_3 +1 , \bar l_{3};P)
-\tilde B (l_3 , \bar l_{3})
\Psi (l_3 , \bar l_{3};P)\,,
\label{3.12a}
\eeq
with
\beq
\label{3.11a}
\tilde A (l_{3}, \bar l_{3})=\frac{
\xi _{0}^{(1 )}(l_{3}+1, \bar l_{3}+1)}
{\xi _{0}^{(1)}(l_{3}, \bar l_{3} +1)}=
-\lambda _{12}
\frac{\tau (l_{1}+1, l_3 +1;\bar l_{3}+1)
\tau (l_{1}, l_3 ;\bar l_{3}+1 )}
{\tau (l_{1}, l_3 +1; \bar l_{3}+1)
\tau (l_{1}+1 , l_3 ; \bar l_{3}+1 )},
\eeq
\beq\label{3.11b}
\tilde B (l_{3}, \bar l_{3})=\frac{
\xi _{0}^{(2)}(l_{3}+1, \bar l_{3})}
{\xi _{0}^{(2)}(l_{3}, \bar l_{3} )}=
\frac{\lambda _{1 2}\lambda _{3 2}}{\lambda _{13}}\,
\frac{\tau (l_{1}+1, l_3 +1;\bar l_{3}+1)
\tau (l_{1}, l_3 ;\bar l_{3} )}
{\tau (l_{1}, l_3 +1; \bar l_{3})
\tau (l_{1}+1 , l_3 ; \bar l_{3}+1 )},
\eeq
is a linear combination of two other equations (\ref{3.10}) written
 in terms of the new variables. 
The  constant prefactors in (\ref{3.11a}), (\ref{3.11b})
are  derived using the reciprocity law for differentials of the
third kind \cite{Fay}.

Alternatively, eqs.\,(\ref{3.12}), (\ref{3.12a}) can be proved 
 in the same way as in Theorem 2.2.

\subsection{Bilinear equations for the $\tau$-function}

We have shown that the Baker-Akhiezer function satisfies an
overdetermined system of linear equations. For compatibility
of this system the coefficient functions must obey certain
non-linear relations. In terms of the $\tau$-function all these
relations have a bilinear form.

\begin{th}
The $\tau$-function obeys the Hirota bilinear difference
equation
\begin{eqnarray}
&&\lambda _{23} \tau (l_1 +1,l_2 , l_3 )
\tau (l_1 , l_2 +1 , l_3 +1 )-
\lambda _{13} \tau (l_1 ,l_2 +1, l_3 )
\tau (l_1 +1, l_2 , l_3 +1 ) \nonumber \\
&+&
\lambda _{12} \tau (l_1 ,l_2 , l_3 +1)
\tau (l_1 +1, l_2 +1 , l_3 )=0
\label{3.24}
\end{eqnarray}
with the
constants $\lambda _{\alpha \beta }$
defined in (\ref{3.17}).
\end{th}
{\it Proof.} First we examine the compatibility  of
eqs.\,(\ref{3.10}) for $(\alpha \beta)=(12)$ and
$(\alpha \beta)=(13)$. The variable $l_1$ is common
for  both linear problems. Eqs.\,(\ref{3.10}) can be
rewritten as
\beq
\Psi (l_{\alpha }+1;P)=M^{(\alpha )}_{1}
\Psi (l_{\alpha };P),\;\;\;\;\;\;\alpha = 2,3,
\label{3.25}
\eeq
where $M^{(\alpha )}_{1}$ is the  difference operator in $l_1$:
\beq
M^{(\alpha )}_{1}=e^{\p _{l_1}}-\lambda _{1\alpha}
\frac{ \tau (l_1 , l_{\alpha})\tau (l_1 +1 , l_{\alpha}+1)}
{ \tau (l_1 , l_{\alpha}+1)\tau (l_1 +1 , l_{\alpha})}.
\label{3.26}
\eeq
Eq.\,(\ref{3.25}) has a family of
linearly independent solutions parametrized by points of the curve
$\G$. Whence the compatibility is equivalent to commutativity
of the operators
\beq
\hat M^{(\alpha )}_{1}=e^{-\p _{l_\alpha }}M^{(\alpha )}_{1},
\label{3.27}
\eeq
i.e.,
\beq
\phantom{a}[\hat M^{(2)}_{1},\;
\hat M^{(3)}_{1}]=0,
\label{3.28}
\eeq
which is a discrete zero curvature condition.
Commuting the operators (\ref{3.26}),
we find after some algebra that this
condition is equivalent to the relation
\begin{eqnarray}
&&\lambda _{13} \tau (l_1 ,l_2 +1, l_3 )
\tau (l_1 +1 ,l_2 , l_3 +1)-
\lambda _{12} \tau ( l_1 , l_2 , l_3 +1)
\tau (l_1 +1 ,l_2 +1 , l_3) \nonumber \\
&+&
H_1 (l_1;l_2 , l_3 ) \tau (l_1 +1 , l_2 , l_3)
\tau (l_1 , l_2 +1 , l_3 +1)=0\,,
\label{3.29}
\end{eqnarray}
where $H_1$ is an arbitrary function such that $H_1 (l_1+1;l_2 , l_3 )=
H_1 (l_1;l_2 , l_3 )$.

All this can be repeated for other two pairs of linear problems,
i.e., for $(\alpha \beta )=(21),\, (23)$ and
$(\alpha \beta )=(31),\, (32)$ in (\ref{3.10}). This leads to
bilinear relations similar to (\ref{3.29}) for the same $\tau$-function
but
with different  functions
 $H_2 $ and
$H_3$. To be
consistent, all three bilinear relations must be identical. This 
determines $H_{1}=-\lambda_{23}$, etc, which proves the theorem.

\noindent
{\bf Remark 2.7}\,\,In the class of  the algebro-geometric solutions
the Hirota equation (\ref{3.24}) is equivalent to Fay's trisecant
identity \cite{Fay}.

The coefficients in eq.\,(\ref{3.24}) may be hidden by the
transformation
\beq
\tau (l_1 , l_2 , l_3 )\rightarrow
\left ( \frac{ \lambda _{13}}{\lambda _{12}}\right )^{l_1 l_2 }
\left ( \frac{ \lambda _{13}}{\lambda _{23}}\right )^{l_2 l_3 }
\tau (l_1 , l_2 , l_3 )
\label{3.30}
\eeq
bringing the Hirota equation into its {\it canonical form}:
\begin{eqnarray}
&&\tau (l_1 +1,l_2 , l_3 )
\tau (l_1 , l_2 +1 , l_3 +1 )-
\tau (l_1 ,l_2 +1, l_3 )
\tau (l_1 +1, l_2 , l_3 +1 ) \nonumber \\
&+&
\tau (l_1 ,l_2 , l_3 +1)
\tau (l_1 +1, l_2 +1 , l_3 )=0.
\label{3.31}
\end{eqnarray}
The formula
\beq
\tau (l_1 , l_2 , l_3 )=\exp \left (
l_1 l_2 \int _{P_3}^{P_2}d\Omega ^{(01)}+
l_2 l_3 \int _{P_1}^{P_2}d\Omega ^{(03)}
\right )
\Theta \left ( \sum _{\alpha =1}^{3}
\vec A(P_{\alpha})l_{\alpha} +\vec Z
\right )
\label{3.32}
\eeq
thus provides the  family of algebro-geometric 
solutions to eq.\,(\ref{3.31}).

This family has $4g+1$ continuous
parameters. Indeed, for $g>1$ the solution depends
on $3g-3$ moduli
of the curve, on $g$ points $\gamma _i$ and
on the 4 marked points $P_{\alpha}$. The dependence
on the choice of local parameters is not essential.

In terms of the variables $l_1,l_3,\bar l_3$ the bilinear 
equations has the form
\begin{eqnarray}
&&\lambda_{13}\tau (l_1 ,l_3 ; \bar l_3 +1 )\tau (l_1 , l_3 +1 ; \bar l_3
)-
\lambda_{23}\tau (l_1 ,l_3 ; \bar l_3 )
\tau (l_1 , l_3 +1; \bar l_3 +1 ) \nonumber \\
&=&\lambda_{12}
\tau (l_1 +1 ,l_3 ; \bar l_3 +1)
\tau (l_1 -1, l_3 +1 ; \bar l_3 ).
\label{3.33}
\end{eqnarray}
(Alternatively, this  equation is a result of the  compatibility of
eqs.\,(\ref{3.10}), (\ref{3.12}) and (\ref{3.12a}).)
In contrast to eq.\,(\ref{3.24}), the variable $l_2$ 
is not shifted and skiped. 
The Hirota equation in the form (\ref{3.24}) can be 
considered as a discrete KP, whereas the form (\ref{3.33})
 is a fully discretized 2D Toda lattice. Let us stress again that 
in the fully discretized setup the KP equation and the 
2D Toda lattice become equivalent.

\subsection{Degenerate cases}

Degenerations of the curve $\G$ lead to important classes
of solutions. 
Among them are multi-soliton
and rational solutions.
Some examples of soliton solutions to
eqs.\,(\ref{3.31}), (\ref{3.33})
were found by R.Hirota \cite{Hirota}. Here we outline the
algebro-geometric construction of  the general
soliton solutions.

Let us concentrate on multi-soliton solutions.
In this case all $N$  ``handles'' of the Riemann surface of genus $N$
become
infinitely thin. In other words, the  algebraic curve of genus $N$
degenerates into 
the complex plane  with a set of $2N$ marked points
$p_i,\, q_i$:
$$
\G\rightarrow \{p_i,\, q_i,\;\; i=1,\ldots ,N\},
$$
where $ p_i,\, q_i$ are the ends of the $i$-th handle. 
The Baker-Akhiezer function
has the same value at each pair $ p_i,\, q_i$.
The punctures $P_{\alpha}$ are replaced by points $z_{\alpha}$
with local parameters $w_{\alpha}=z-z_{\alpha}$.
In this case the meromorphic differentials of the third kind have the
form:
$$
d\Omega ^{(\alpha \beta )}=
\frac{(z_{\beta }-z_{\alpha })dz}
{(z-z_{\beta })
(z-z_{\alpha })}.
$$
Let $F_0(z)$ be a polynomial of degree $N$:
$$
F_0(z)=z^N +
\sum _{j=1}^{N}b_j z^{N-j}=\prod_{i=1}^{N}(z-\gamma _{i}).
$$
Its zeros $\gamma_i$ will stand for poles of
the Baker-Akhiezer function.

Let us concentrate on the particular case when there
are four punctures and three cuts from $z_0$ to $z_{\alpha}$
($\alpha = 1,2,3$) on the complex  $z$-plane. Here $z_{\alpha}$ are
the $z$-coordinates of the marked points $P_{\alpha}$: 
$z_{\alpha}=z(P_{\alpha })$. In this case the general 
definition of the 
Baker-Akhiezer function  suggests the Ansatz:
\beq
\Psi (l;z)=
\frac{F(l;z)}{F_0 (z)}\Psi _{0}(l;z)
\label{s1}
\eeq
with  
\beq
\Psi _{0}(l;z)=
\prod _{\alpha =1}^{3}\left (\frac{z-z_{\alpha }}
{z-z_{0}}\right )^{l_{\alpha}},\;\;\;\;\;\;l\equiv (l_1 , l_2 , l_3 )
\label{germ}
\eeq
and the polynomial
\beq
F(l;z)=\sum _{j=0}^{N}a_j (l)z^{N-j}
\label{s2}
\eeq
 with yet undetermined $l$-dependent coefficients.
These coefficients
are determined by  $N$ conditions
\beq
\Psi (l;p_i )=\Psi (l; q_i ), \;\;\;\;\;\;
l=1, \ldots , N.
\label{s3}
\eeq

They  
are equivalent to the
 system of $N$ linear equations for
$N+1$ unknown coefficients $a_j$ in eq.\,(\ref{s2}):
$$
\sum_{j=0}^{N}
K_{ij}a_j=0, \;\;\;\;i=1,\ldots , N, \;\;\;\;
j=0,1,\ldots , N,
$$
where
\beq
K_{ij}=\frac{p_i ^{N-j}}{F_0 (p_i )}
\Psi _{0}(l;p_i )
-\frac{q_i ^{N-j}}{F_0 (q_i )}
\Psi _{0}(l;q_i ).
\label{M}
\eeq
Solving the system of linear equations 
 we represent the Baker-Akhiezer function in the form
\beq
\Psi (l;z)=c(l){\Delta (l;z)\over \Delta (0;z)}
\Psi _{0}(l;z),
\label{baker}
\eeq
where
$$
\Delta (l;z)=
\det_{ij}(\hat K_{ij})
$$
and $\hat K$ is the $(N+1)\times (N+1)$-matrix with entries
$$
\hat K_{0j}=z^{N-j},\;\;\;\;
\hat K_{ij}=K_{ij},\;\;\;\;
i=1, \ldots ,N.
$$
The normalization factor $c(l)$ is fixed by the asymptotics
$$
\Psi (l;z)=(z-z_{0})^{-l_1 -l_2 -l_3 }(1+O(z-z_{0}))
$$
near the point $z_0$. It gives
$$
c(l)=\frac{\Delta (0;z_{0})}{\Delta (l;z_0 )}
\prod _{\alpha =1}^{3}(z_0 -z_{\alpha })^{-l_{\alpha }}.
$$
We also point out the identity
\beq
\Delta (l_{\alpha};z_{\alpha})=
\Delta (l_{\alpha}+1;z_{0}), \;\;\;\;\;\; \alpha =1,2,3.
\label{s4}
\eeq

The degeneration of the $\tau$-function (\ref{tau}) is
\beq
\tau (l)=\Delta (l; z_0 ).
\label{trigtau}
\eeq
It satisfies the bilinear equation (\ref{3.24}) with
$$
\lambda _{\alpha \beta}=\frac{ z_{\alpha}-z_{\beta}}
{(z_0 -z_{\alpha})
(z_0 -z_{\beta})}.
$$
The continuous parameters of the solution (\ref{trigtau}) are
$2N$ points $p_i, q_i$, $N$ points $\gamma _{i}$ and 4 points
$z_{\alpha}$. However, the $\tau$-function is invariant (up to
an irrelevant constant) under the
simultaneous fractional-linear transformation
$z\rightarrow (az+b)/(cz+d)$,
$ad-bc=1,$ of all these parameters, so we are left with
$3N+1$ parameters.

In the case of rational degeneration the points $p_i$ and $q_i$ merge and
 the condition (\ref{s3})
becomes
\beq
({\hat {\cal D}}_{i} \Psi )(p_i )=0,\;\;\;\;\;\;\;\;
i=1,\ldots , N,
\label{rational}
\eeq
where
${\hat {\cal D}}_{i}$ are some differential operators in $z$
with constant coefficients. This changes the form of the matrix
$K$ but  eqs.\,(\ref{baker}), ({\ref{s4}) remain the same.

\section{Elliptic solutions}

General finite-gap solutions of
Hirota's equation for  an arbitrary algebraic curve
$\G$ with  punctures $P_{\alpha}$ are 
{\it quasi-periodic} functions
of all  variables $l_\alpha$. Below we  construct a special important
class of 
 solutions for which the quantity (\ref{I5}) is {\it doubly periodic} 
 in one of the variables. The $\tau$-function in this case is an elliptic 
 polynomial (\ref{I4}). We call them elliptic solutions.  
 We show that the elliptic solutions also imply a spectral
algebraic curve and are therefore a subclass of the algebro-geometric
 solutions of Sect.2.

Among all algebro-geometric solutions described in the previous
section the elliptic solutions in one of the variables or their linear 
combination are characterized as follows. 
Let  
\beq \label{vf}
\p_{x} = \sum X_{\alpha \beta }\p_{l_{\alpha \beta }}
\eeq
be a vector field in the space of variables $l_{\alpha \beta}$ and let
$\vec U =\sum _{\alpha \beta }
X_{\alpha \beta}\vec U^{(\alpha \beta )}$. Let us transport the
 $\tau$-function (\ref{F}) along the vector field $\p_{x}$ and denote it
by
\beq
\label{F1}\tau(x)=\Theta \left (\vec U x+\vec X(l)+\vec Z\right).
\eeq
Consider a set of algebraic
curves with punctures $P_{\alpha}$ and cuts such that the vector
$\vec U$ has the property:
\begin{itemize}
\item{There exist two constants
$\omega _{1},\omega _{2}$, 
${\rm Im} \  (\omega _{2}/ \omega _{1}) \neq 0$, such that
$
2\omega _{a}\vec U $, $a=1,2$, belong to the
lattice of periods of
the holomorphic differentials on $\G$:
$$\tau(x+2\omega_a)=e^{r_ax+s_a}\tau(x)\,,$$
where  $r_a,s_a$ are constants.}
\end{itemize}

\noindent
The  $\tau$-function is  then an
elliptic polynomial
in the variable $x$. 
Due to the commensurability of $\vec U$ and the lattice of periods,
 solutions to the  equation 
\beq
\Theta (\vec U x+\vec Z)=0
\label{e2}
\eeq
are
$x_i (\vec Z) +2J_1 \omega _{1}+2J_2 \omega _{2}$, where 
$x_i (\vec Z)$ belong to
the fundamental domain of the lattice generated by $2\omega _{1}$, 
$2\omega _{2}$ and $J_1$, $J_2$ run over all integers.
Therefore, 
\beq
\Theta (\vec U x+\vec Z)
=e^{a_1 x +a_2 x^{2}}\prod_{i=1}^{N}
\sigma\big (x-x_i (\vec Z)\big )
\label{e3}
\eeq
with  $x$-independent $a_1 , a_2$ and $N\geq g$.

The requirement of the  ellipticity imposes
$2g$ constraints on $4g+1$ parameters. Therefore the dimension of
 the family of elliptic solutions
 is  $2g+1$.

Below we concentrate on a  specific case where the ellipticity is imposed
along
$l_{01}\equiv l_1$ by setting 
$\vec U = \eta ^{-1} \vec U^{(01)}$ and $x=\eta l_1$  where $\eta$ is a
complex constant.

The logic of this section is opposite to the one of Sect.\,2. 
Here we solve the {\it direct problem} and 
show that {\it all} solutions which are elliptic 
in any "direction" $l_{\alpha\beta}$
of the form (\ref{I4}) with
a time-independent degree $N$,  
are of the finite gap type (\ref{F1}).

\subsection{Equations of motion for zeros of the $\tau$-function}

Let us show how to obtain the equations of motion for zeros of
elliptic solutions by elementary methods.

First of all, let us  rename variables to emphasize the "direction" of
ellipticity:
\beq
l_1\equiv \eta x,\;\;\;l_2\equiv n,\;\;\;l_3=m,\;\;\;\bar l_3\equiv \bar
m,
\label{not}
\eeq
so that the $\tau$-function has the form (\ref{I4}). Let us now consider
one  
of eqs.\,(\ref{3.10}), say for $\alpha,\beta=1,2$. In this equation
 all variables except $x$ and $m$ are parameters and we skip them wherever 
it does not cause confusion:
$x_i^{m,n}\rightarrow x^{m}_{i},\;\tau ^{m,n}(x)\rightarrow \tau ^{m}(x)$
\beq
\psi ^{m}(x+\eta )-\lambda_{13}
\frac{\tau ^{m}(x)\tau ^{m+1}(x+\eta )}
{\tau ^{m+1}(x)\tau ^{m}(x+\eta )}
\psi ^{m}(x)
=\psi ^{m+1}(x)\,.
\label{e5}
\eeq
Let us look for solutions of the form
$$
\psi ^{m}(x)=\frac{\rho ^{m}(x)}{\tau ^{m}(x)}
$$
with some  function $\rho ^{m}(x)$. 
Eq.\,(\ref{e5}) then
reads 
\beq
\tau ^{m+1}(x)\rho ^{m}(x+\eta )
-\lambda_{13}\tau ^{m+1}(x+\eta )\rho ^{m}(x)
=\tau ^{m}(x+\eta )\rho ^{m+1}(x).
\label{e6}
\eeq
We are interested in the case when $\tau ^{m}(x)$ is an
elliptic polynomial in $x$ for any $m$:
\beq
\tau ^{m}(x)=\prod _{j=1}^{N}\sigma (x-x_{j}^{m}).
\label{e7}
\eeq
The "equations of motion" for its roots $x_{j}^{m}$ in the
"discrete time" $m$ can be easily obtained from eq.\,(\ref{e6})
in the following way. Substituting $x=x_{j}^{m+1}$,
$x=x_{j}^{m+1}-\eta $,
$x=x_{j}^{m}-\eta $, we get the relations
\begin{eqnarray}
&&-\lambda_{13}\tau ^{m+1}(x_{j}^{m+1}+\eta )\rho ^{m}(x_{j}^{m+1})
=\tau ^{m}(x_{j}^{m+1}+\eta )
\rho ^{m+1}(x_{j}^{m+1}),
\nonumber \\
&&\tau ^{m+1}(x_{j}^{m+1}-\eta )\rho ^{m}(x_{j}^{m+1})
=\tau ^{m}(x_{j}^{m+1})
\rho ^{m+1}(x_{j}^{m+1}-\eta ),
\nonumber \\
&&\tau ^{m+1}(x_{j}^{m}-\eta )\rho ^{m}(x_{j}^{m})
=\lambda_{13} \tau ^{m+1}(x_{j}^{m})
\rho ^{m}(x_{j}^{m}-\eta ),
\label{e8}
\end{eqnarray}
respectively.
Combining these relations, we eliminate $\rho$
and obtain a system of equations for the roots $x_i^m$:
\beq
\prod _{k=1}^{N}
\frac{
\sigma (x_{j}^{m}-x_{k}^{m-1})
\sigma (x_{j}^{m}-x_{k}^{m}+\eta )
\sigma (x_{j}^{m}-x_{k}^{m+1}-\eta)}
{\sigma (x_{j}^{m}-x_{k}^{m-1}+\eta )
\sigma (x_{j}^{m}-x_{k}^{m}-\eta )
\sigma (x_{j}^{m}-x_{k}^{m+1})}
=-1.
\label{e9a}
\eeq

Note that these equations involve only one discrete variable $m$
while the direct substitution of the Ansatz (\ref{e7}) into
the non-linear equation (\ref{3.31}) would give a system of
equations involving two discrete variables. It is not easy to
see that they in fact decouple. The decoupling becomes transparent
if one starts with the auxiliary linear problem (\ref{e5}).

Similar equations hold  for the discrete $l_3\equiv\bar m$-dynamics.
From the linear problem (\ref{3.12}) we obtain the equations for the 
$\bar m$ dependence of $x_i$:
\beq
\prod _{k=1}^{N}
\frac{
\sigma (x_{j}^{\bar m}-x_{k}^{\bar m-1}-\eta )
\sigma (x_{j}^{\bar m}-x_{k}^{\bar m}+\eta )
\sigma (x_{j}^{\bar m}-x_{k}^{\bar m+1})}
{\sigma (x_{j}^{\bar m}-x_{k}^{\bar m-1} )
\sigma (x_{j}^{\bar m}-x_{k}^{\bar m}-\eta )
\sigma (x_{j}^{\bar m}-x_{k}^{\bar m+1}+\eta )}
=-1.
\label{e9b}
\eeq
To avoid confusion, let us stress that the $x_i$ depend on all discrete
times.
However, the variables are separated in the equations of motions.

{\bf Remark 3.1}\,\,Another choice of the "direction" of ellipticity gives
rise to
different equations of motion. To illustrate this, let us require 
$\tau (l)$ to be an elliptic polynomial
in the direction orthogonal to $s=l_3+\bar l_3$ and $a=l_1+l_3-\bar l_3$, 
so that the zeros of the $\tau$-function depend on $a$ and $s$. Then the
vector 
field in (\ref{vf}) is
$\partial_ x=\eta\partial_{l_1}-
\frac{1}{2}\partial_{l_3}+\frac{1}{2}\partial_{\bar l_3}$.
In terms of
$T^{a,s}(x)\equiv \tau (l_1 , l_3,\bar l_3 )$ bilinear
equation (\ref{3.33}) reads:
$$
\lambda _{12}T^{a,s}(x+\eta )T^{a,s}(x-\eta )
+\lambda _{23}T^{a,s+1}(x)T^{a, s-1}(x)
= \lambda _{13}T^{a+1,s}(x)T^{a-1,s}(x)
$$
and the zeros $x_{i}^{a,s}$ of  $T^{a,s}(x)$ obey the system of
coupled equations:
$$
\frac{T^{a+1, s+1}(x_{i}^{a,s})
T^{a-1, s}(x_{i}^{a,s}-\eta )
T^{a, s-1}(x_{i}^{a,s}+\eta )}
{T^{a-1, s-1}(x_{i}^{a,s})
T^{a+1, s}(x_{i}^{a,s}+\eta )
T^{a, s+1}(x_{i}^{a,s}-\eta )}=-1\,,
$$
$$
\frac{T^{a+1, s-1}(x_{i}^{a,s})
T^{a-1, s}(x_{i}^{a,s}-\eta )
T^{a, s+1}(x_{i}^{a,s}+\eta )}
{T^{a-1, s+1}(x_{i}^{a,s})
T^{a+1, s}(x_{i}^{a,s}+\eta )
T^{a, s-1}(x_{i}^{a,s}-\eta )}=-1\,,
$$
These equations are more complicated than (\ref{e9a}). In
 contrast to the previous case, evolutions in $a$ and $s$ are not
separated.

\subsection{Double-Bloch solutions to the linear problems}

In order to further examine the elliptic solutions, we need the
notion of double-Bloch functions.
A meromorphic function $f(x)$ is said to be {\it double-Bloch} if it
enjoys the following monodromy properties:
\beq
f(x+2\omega_{a})=B_{a} f(x), \;\;\;\;\;\;\;  a=1,2.
\label{DB1}
\eeq
The
complex numbers $B_{a}$ are called {\it Bloch
multipliers}. A
non-trivial double-Bloch function
can be
represented as a linear combination of elementary ones:
\beq
f(x)=\sum_{i=1}^N c_i\Phi(x-x_i,\zeta) k ^{x/\eta},
\label{DB2}
\eeq
where \cite{kz}
\beq
\Phi(x,\zeta )=
{\sigma(\zeta +x+\eta)\over \sigma(\zeta +\eta) \sigma(x)} \left[
{\sigma(\zeta -\eta)\over \sigma(\zeta +\eta)}\right]^{x/(2\eta )}
\label{DB3}
\eeq
and complex parameters $\zeta$ and $k$ are related to the
Bloch multipliers by the formulas
\beq
B_{a}=k^{2\omega _{a}/\eta}
\exp (2\zeta (\omega _{a})(\zeta +\eta ))
\left (\frac{\sigma (\zeta -\eta )}{\sigma
(\zeta +\eta )}\right )^{\omega_{a}/\eta}
\label{DB4}
\eeq
($\zeta (x)=\sigma '(x)/\sigma (x)$
is the Weierstrass $\zeta$-function).

Let us point out some properties of the function $\Phi (x, \zeta )$.
Considered as a function of $\zeta$,
$\Phi(x,\zeta )$ is double-periodic:
$$
\Phi(x,\zeta +2\omega_{a})=\Phi(x,\zeta ).
$$
For general values of $x$ one can define a single-valued branch of
$\Phi(x,\zeta )$ by cutting the elliptic curve between the points
$\zeta =\pm \eta$.
In the fundamental domain of the lattice  generated by
$2\omega_{a}$ the function $\Phi(x,\zeta )$ has a unique
pole at the point $x=0$:
$$
\Phi(x,\zeta )={1\over x}+O(1)\,.
$$
In the next subsection we need the identity:
\beq
\Phi (x,z)\Phi (y,z) =\Phi (x+y ,z)
\big ( \zeta (x) +\zeta (y) +\zeta (z+\eta )-
\zeta (x+y+z+\eta )\big )
\label{id}
\eeq
which is equivalent to the well known 3-term bilinear
functional equation for the $\sigma$-function.

Recall the notion of equivalent Bloch multipliers \cite{kz}.
The "gauge transformation"
$f(x)\rightarrow \tilde f(x)=f(x)e^{bx}$
($b$ is an arbitrary constant) does not change the poles of any
function and transforms a double-Bloch function into
another double-Bloch function. If $B_{a}$ are Bloch multipliers for
$f$, then the Bloch multipliers for $\tilde f$ are
$\tilde B_1=B_1e^{2b\omega_1}$, $\tilde B_2=B_2 e^{2b\omega_2}$.
Two such pairs of Bloch multipliers $B_a$ and $\tilde B_a$ are said to be
{\it equivalent}. 
 (In other words, they are equivalent
if the product $B_{1}^{\omega_2} B_{2}^{-\omega_1}$ is the same for
both pairs.)

This definition implies that any double-Bloch function
can be represented as a ratio of two elliptic polynomials of the
same degree multiplied by an exponential function and a constant:
\beq
f(x)=c' (k')^{x/\eta }\prod _{i=1}^{N}
\frac{ \sigma (x-y_{i})}
{\sigma (x-x_{i})}.
\label{DB5}
\eeq
The Bloch multipliers are
$$
B_a = (k')^{2\omega _{a}/\eta}\exp \left (2\zeta (\omega _{a})
\sum _{j=1}^{N}(x_j - y_j )\right ).
$$
Eqs.\,(\ref{DB2}) represents a Bloch function by its poles and residues,
whereas
eq.\,(\ref{DB5}) represents a Bloch function by its poles and zeros.

\subsection{The Lax representation}

The coefficients in eq.\,(\ref{e5}) are elliptic functions, i.e.
double-periodic with periods $2\omega _{a}$. Therefore, the equation has
double-Bloch
 solutions .
Similarly to the case
of the Calogero-Moser model and its spin generalizations
the dynamics of poles of the
elliptic coefficient in the linear problem
is determined by the fact that
equation (\ref{e5}) has an infinite number of
double-Bloch solutions.

In what follows we always assume that
the poles are in a generic position, i.e.
$x^{m}_{i}-x^{m}_{j}\neq 0, \pm \eta$ and
$x^{m}_{i}-x^{m \pm 1}_{j}\neq 0, \pm \eta$ for any pair
$i\neq j$. Exceptional cases are also of interest but
must be treated separately.

\begin{th} Let $\tau ^{m}(x)$ be an elliptic polynomial of degree
$N$ .
Equation (\ref{e5})
has $N$ linearly independent double-Bloch solutions
with simple poles at the points
$x_i^{m}$ and equivalent Bloch multipliers
if and only if
zeros $x_i^{m}$ of the $\tau$-function
satisfy "equations of motion" (\ref{e9a}).
\end{th}
\begin{th}
If eq.\,(\ref{e5}) has $N$ linearly independent double-Bloch
solutions with equivalent Bloch multipliers, then it has
infinite number of them.
All these solutions
have the form
\beq
\psi ^{m}(x)=\sum_{i=1}^N c_i(m,\zeta , k)
\Phi(x-x_i^{m},\zeta ) k^{x/\eta}
\label{e10}
\eeq
($\Phi (x,\zeta )$ is defined in (\ref{DB3})).
The set of corresponding pairs $(\zeta ,z)$ is parametrized by points
of an algebraic curve.
\end{th}

These theorems are proved by the same arguments as in \cite{kz}. 
Here we present the main steps.

$N$ linearly independent double-Bloch
solutions with equivalent Bloch multipliers
may be written in the form (\ref{e10}) with some values
of the parameters $\zeta _{r}, k_r,\; s=1,\ldots, N$.
Equivalence of the multipliers implies that the $\zeta _{r}$ can
be chosen to be equal
$\zeta _{r}=\zeta.$

Let us substitute the function $\psi ^{m}(x)$ of the form (\ref{e10})
with this particular value of $\zeta$ into eq.\,(\ref{e5}).
Since any double-Bloch function (except equivalent to a constant) has at
least one pole,
it follows that the equation is satisfied if its left hand side  has zero
residues at the points $x=x_i^m- \eta$ and $x=x_i^{m+1}$.
The cancelation of poles at these points
gives the conditions
\beq
k c_i(m,\zeta ,k)
-f_{i}(m) \sum_{j=1}^{N}
c_j(m,\zeta ,k)\Phi(x_i^m -x_j^m -\eta ,\zeta)=0\,,
\label{e11}
\eeq
\beq
c_i(m+1,\zeta ,k)=g_{i}(m) \sum_{j=1}^{N}
c_j(m,\zeta ,k) \Phi(x_i^{m+1}-x_j^m ,\zeta )\,,
\label{e12}
\eeq
where
\beq
f_{i}(m)=\lambda_{13}\frac{
\prod_{s=1}^{N}
\sigma(x_i^m- x_s^{m}-\eta )\sigma(x_i^m - x_s^{m+1})}
{\prod _{s=1, \ne i}^{N}\sigma(x_i^m - x_s^{m})
\prod _{s=1}^{N}\sigma(x_i^m - x_s^{m+1}-\eta )}\,,
\label{e13}
\eeq
\beq
g_{i}(m)=-\lambda_{13}\frac{
\prod_{s=1}^{N}
\sigma(x_i^{m+1}- x_s^{m+1}+\eta )\sigma(x_i^{m+1} - x_s^{m})}
{\prod _{s=1, \ne i}^{N}\sigma(x_i^{m+1} - x_s^{m+1})
\prod _{s=1}^{N}\sigma(x_i^{m+1} - x_s^{m}+\eta )}\,.
\label{e14}
\eeq

Introducing a vector $C(m)$ with components
$c_i(m, \zeta , z)$ we can
rewrite these
conditions in the form
\beq
({\cal L}(m)-kI)C(m)=0\,,
\label{e15}
\eeq
\beq
C(m+1)={\cal M}(m)C(m)\,,
\label{e16}
\eeq
where $I$ is the unit matrix. The matrix elements of
 ${\cal L}(m)$ and ${\cal M}(m)$ are:
\beq
{\cal L}_{ij}(m)=f_{i}(m) \Phi
(x_i^{m} - x_j^m -\eta ,\zeta ),
\label{Lax}
\eeq
\beq
{\cal M}_{ij}(m)=g_{i}(m)\Phi(x_i^{m+1}- x_j^m , \zeta ).
\label{Max}
\eeq
The compatibility condition of (\ref{e11}) and
(\ref{e12}),
\beq
{\cal L}(m+1){\cal M}(m)={\cal M}(m){\cal L}(m)
\label{e17}
\eeq
has a form of the discrete Lax
equation. The Lax equation (\ref{e17}) appeared 
in ref.\,\cite{NRK}, where eqs. (\ref{e9a}) were proposed
as a time discretezation of the RS model.

\begin{lem}
For matrices ${\cal L}$ and ${\cal M}$
defined by (\ref{Lax}),
(\ref{Max})
the discrete Lax equation (\ref{e17}) is equivalent to the  equations
(\ref{e9a}).
\end{lem}
The proof is along the lines of ref.\,\cite{NRK}.
 We have
\begin{eqnarray}
&&F_{ij}\equiv(x, \zeta )({\cal M}(m){\cal L}(m)-
{\cal L}(m+1){\cal M}(m))_{ij}
\nonumber \\
&=&f_{i}(m+1)\sum _{s}
g_{s}(m)\Phi(x_i^{m+1}-x_s^{m+1}-\eta ,\zeta )
\Phi(x_s^{m+1}-x_j^m ,\zeta )\nonumber \\
&+&g_i (m)\sum_s f_s (m)\Phi(x_i^{m+1}-x_s^m,\zeta )
\Phi(x_s^m-x_j^m -\eta,\zeta ).
\label{e18}
\end{eqnarray}
The coefficient in front of the
leading singularity at $\zeta =-\eta$ is proportional to
$$
f_i(m+1)\sum_s g_s(m) + g_i(m) \sum_s f_s (m).
$$
On the other hand,
\beq
\sum _{s}\big ( f_s (m)-g_s (m)\big )=0
\label{e19}
\eeq
(because this is the sum of residues of the elliptic coefficient
in eq.\,(\ref{e5}) ).
Therefore,
\beq
f_i(m+1)=g_i(m),\;\;\;\;\;\;\; i=1, \ldots , N.
\label{e20}
\eeq
These equations are equivalent to (\ref{e9a}).

To show that (\ref{e18})
is identically zero provided eqs.\,(\ref{e20}) hold,
we use the identity (\ref{id}):
$$
F_{ij}(x, \zeta )=-g_{i}(m)\Phi (x_{i}^{m+1}-x_{j}^{m}-\eta ,
\zeta )\,G,
$$
\begin{eqnarray}
G&=&
-\sum_s f_s(m)\big (\zeta(x_s^{m+1}-x_s^m)+\zeta(x_s^m-x_j^m)-\eta )
\big )\nonumber \\
&+&\sum_s
g_s(m)
(\zeta(x_i^{m+1}-x_s^{m+1}-\eta )+\zeta(x_s^{m+1}-x_j^m)\big ).
\label{G}
\end{eqnarray}
Noting that $G$ is proportional to the sum of residues of the
elliptic function
$$
\big [\zeta (x_i^{m+1}-\eta -x)+\zeta(x-x_j^{m})\big ]
\prod _{i=1}^{N}\frac{
\sigma (x-x_{i}^{m})
\sigma (x-x_{i}^{m+1}+\eta )}
{\sigma (x-x_{i}^{m}+\eta )
\sigma (x-x_{i}^{m+1})}
$$
at the points $x=x_i^{m+1}$ and $x=x_i^m-\eta$, we conclude that $G=0$.

It was already proved that eq.\,(\ref{e5}) has $N$ linearly
independent solutions if eqs.\,(\ref{e9a}) or the Lax equation
(\ref{e17}) hold for some value of the spectral parameter $\zeta$.
It then follows from Lemma 3.1 that the Lax equation holds for
any value of $\zeta$. Therefore, for each $\zeta$ there exists
a double-Bloch solution given by (\ref{e10}), where the $c_i$ are
components of the common solution to (\ref{e15}), (\ref{e16}).

\begin{th}
All elliptic solutions of eq.\,(\ref{3.24}) of the form
(\ref{e7}) are of the algebro-geometric type and $x_{i}^{m}$
are given implicitly by the equation
\beq
\Theta \left ( \eta ^{-1}\vec A (P_1 )x_{i}^{m}
+m\vec A (P_3 ) +\vec Z \right ) =0,
\label{e21}
\eeq
where the Riemann theta-function corresponds to the algebraic
curve $\G$ defined by the characteristic equation
\beq
R(k ,\zeta )\equiv \det ({\cal L}(m)+k I)
=k^N +\sum _{i=1}^{N}r_{i}(\zeta )k^{N-i}=0\,.
\label{curve}
\eeq
The matrix ${\cal L}$ is given by
eqs.\,(\ref{Lax}), (\ref{e13})
and the coefficients $r_{i}(\zeta )$ have the form
$$
r_{i}(\zeta )=\frac{\sigma (\zeta +\eta )^{(i-2)/2}
\sigma (\zeta -(i-1)\eta )}
{\sigma (\zeta -\eta )^{i/2}}I_{i}
$$
where $I_i$
are integrals of motion. The characteristic equation (\ref{curve})
and $I_i$ are functions of $x_{i}^{m }$ and $x_{i}^{m+1}$
but stay the same for all $m$.
The spectral curve $\G$ determined by eq.\,(\ref{curve}) 
is an algebraic curve
realized as a ramified
covering of the elliptic curve. The function (\ref{e10})
is the Baker-Akhiezer function on $\G$.
\end{th}
We call the spectral curve $\G$ defined in Theorem 3.3 the
Ruijsenaars-Schneider (RS) curve.
The RS curve is identical to the
spectral curve for the continuous time RS model
studied in ref.\,\cite{kz}. The proof of the theorem 
is omitted. It is as in ref.\,\cite{kz}.

The matrix ${\cal L}$ is defined by fixing $x_{i}^{m_{0}}$ and
$x_{i}^{m_{0}+1}$. These Cauchy data uniquely define the RS
curve $\G$, the vectors
$\vec A (P_{\alpha })$, $\alpha =1,2$ and
$\vec Z$ in eq.\,(\ref{e21}). The curve and the vectors
$\vec A (P_{\alpha })$, $\alpha =1,2$ do not depend on the choice
of $m_0$.
They are action-type variables.
The vector $\vec Z$ depends linearly on this
choice and its components are thus angle-type variables.

\noindent
{\bf Remark 3.2}\,\,The discrete time dynamics defined in Theorem 3.3
is time-reversible, i.e. the Cauchy data
$x_{i}^{m_{0}}$, $x_{i}^{m_{0}+1}$ completely determine
the dynamics for both
time directions
up to permutation of the "particles".
The ${\cal L}$-${\cal M}$-pair for the
backward time motion is obtained from the difference
equations for the dual Baker-Akhiezer function (see Remark 2.6)
with an Ansatz similar to (\ref{e10}).
An alternative way to derive equations of motion (\ref{e9a})
is to require 
the spectral equation
(\ref{e15}) to be identical to the similar equation for
the backward time motion.

\noindent
{\bf Remark 3.3}\,\,The form of equations for the dynamics in $l_2\equiv
n$
 is identical to the equations (\ref{e9a}) of the dynamics in $m\equiv
l_3$. 
The Cauchy data for $m$-dynamics, i.e., values of $x_i^m$ at $m=0$ and
$m=1$  
completely determine an
 evolution and Cauchy data 
in $n$ (as well as all other  flows). Comparing  ${\cal L}$-operators for
each flow,
one finds relations between the Cauchy data:
\beq
\prod _{s=1}^{N}\frac{
\sigma (x_{i}^{0,0}-x_{s}^{1,0})
\sigma (x_{i}^{0,0}- x_{s}^{0,1}-\eta )}
{\sigma (x_{i}^{0,0}- x_{s}^{0,1})
\sigma (x_{i}^{0,0}-x_{s}^{1,0}-\eta )}=
\frac{ \lambda_{12} }{\lambda_{13} }, \;\;\;\;\;\;\;i=1,2, \ldots , N.
\label{e22}
\eeq
Similar connections  exist for the initial data
of  the $\bar m$-flow.

\subsection{Loci equations}

The values of $x_{i}^{0}$, $x_{i}^{1}$ may be arbitrary if
no other reduction (apart from the elliptic one) is imposed.
If there is an additional reduction, then the 
$x_{i}^{0}$, $x_{i}^{1}$ are constrained to belong to a
submanifold of ${\bf C}^{2N}$, the {\it reduction locus}.
An example of such a locus in the continuous setup is the
KP\,$\rightarrow$\,KdV locus (\ref{locus}). Here we present
equations defining the loci for three important reductions
of Hirota's difference equation. In these cases spectral 
curves of algebro-geometric solutions are hyperelliptic. As before,
$x_{i}^{0}$, $x_{i}^{1}$ are assumed to be in generic position.

{\it A) Discrete KdV equation \cite{HirotaKdV}.}
Discrete KdV equation appears as 
the reduction 
$$\tau (l_{1},l_2 +1, l_3 +1)=\tau (l_{1},l_2 , l_3 ), \;\;\;\;\;\;\;
x\equiv \eta l_1
$$
of the general 3-dimensional Hirota equation (\ref{3.24}). 
In the notation (\ref{not}) the equation is
\beq\label{KDV}
\lambda_{23}\tau^m(x+\eta)\tau^m(x)-\lambda_{13}\tau^{m-1}(x)\tau^{m+1}(x+\eta)
+\lambda_{12}\tau^{m+1}(x)\tau^{m-1}(x+\eta)\,.\eeq
For algebro-geometric solutions the reduction 
means that $\vec U^{(02)}+\vec U^{(03)}$
belongs to the lattice of periods. Therefore, the function
\beq
\varepsilon (P)=\exp \left ( \int _{Q_0 }^{P}( d\Omega ^{(02)}
+d\Omega ^{(03)})\right )
\label{E}
\eeq
is meromorphic on the curve $\G$. From the definition of
the abelian integrals it follows that this function has
a double pole at $P_0$ and simple zeros at $P_2$  and
$P_3$. Outside these points $\varepsilon (P)$ is holomorphic and
is not  zero. The existence of such a function means that the spectral
curve is
 {\it hyperelliptic}. A hyperelliptic  curve of genus $g$ 
 can be defined by the equation
\beq
y^2 =\prod _{i=1}^{2g+1} (\varepsilon -\varepsilon _{i}).
\label{hyper}
\eeq

This is a two-fold covering of the complex plane of the
variable $\varepsilon$.
The projection of $\G$ onto the $\varepsilon$-plane defines
$\varepsilon$ as a meromorphic
function on $\G$. This function has a
double pole on $\G$ at the branch point $P_{\infty}$
(above $\varepsilon =\infty$) and two
simple zeros at the points $P_0^{(\pm )}$
(above $\varepsilon =0$).
The identification of this notation for the punctures
with our previous ones is
$$
P_{\infty }=P_0,\;\; P_0^{(-)}=P_2,\;\; P_0^{(+)}=P_3.
$$

The branch points $\varepsilon _{i}$ in (\ref{hyper}) may not
be arbitrary since the curve should simultaneously be of the
RS type.
Correspondingly, the Cauchy data
$x_{i}^{0}$, $x_{i}^{1}$
with respect to the $l_3$-flow
obey certain constraints. Using the equations of motion (\ref{e9a}) and
 (\ref{e22}), we  obtain a system of $2N$ coupled equations
on allowed values of $x_{i}^{0}$, $x_{i}^{1}$ ("equilibrium locus"):
\beq
\prod _{s=1}^{N}\frac{
\sigma (x_{i}^{1}-x_{s}^{1}+\eta )
\sigma (x_{i}^{1}-x_{s}^{0}-\eta )}
{\sigma (x_{i}^{1}-x_{s}^{1}-\eta )
\sigma (x_{i}^{1}-x_{s}^{0}+\eta )}
=-\, \frac{\lambda _{13}}{\lambda _{12}},
\label{l1a}
\eeq
\beq
\prod _{s=1}^{N}\frac{
\sigma (x_{i}^{0}-x_{s}^{0}+\eta )
\sigma (x_{i}^{0}-x_{s}^{1}-\eta )}
{\sigma (x_{i}^{0}-x_{s}^{0}-\eta )
\sigma (x_{i}^{0}-x_{s}^{1}+\eta )}
=-\, \frac{\lambda _{12}}{\lambda _{13}}
\label{l1b}
\eeq
for $i=1,2, \ldots , N$.  With the help of eq.\,(\ref{3.17a}) the
right hand sides can be expressed through abelian integrals.
The relation between the number of zeros $N$
and genus $g$ of the spectral curve is a subtle question.
 We do not discuss it here.

Each of the systems (\ref{l1a}),
(\ref{l1b}) has the form of Bethe equations for an $N$-site
spin chain of spin 1 at each site. One may treat $x_{s}^{0}$
(for instance)  as arbitrary input parameters
while Bethe's quasimomenta $x_{s}^{1}$ are to be determined by
eqs.\,(\ref{l1a}). However,  the  system
of locus equations (\ref{l1a}),\,(\ref{l1b}) determines
 $x_{i}^{0}$ and $x_{i}^{1}$ simultaneously.

In the continuous time limit we set $x_{i}^{0}=x_i$,
$x_{i}^{1}=x_i +\epsilon \dot x_{i} +
\frac{1}{2}\epsilon ^{2} \ddot x_{i}$,
$\dot x_{i}=\p _{t}x_{i}$, $\epsilon \rightarrow 0$.
Assuming $\lambda _{13}/\lambda _{12}\rightarrow -(1+C\epsilon )$
with a constant $C$, we get from (\ref{l1a}), (\ref{l1b}):
\beq
\sum _{k=1}^{N}\dot x_{k}
\left [ \zeta (x_i - x_k -\eta ) -\zeta (x_i -x_k +\eta )\right ]
=C,
\label{l2a}
\eeq
\beq
\sum _{k=1}^{N}\dot x_{k}
\left [ \wp (x_i - x_k -\eta ) -\wp (x_i -x_k +\eta )\right ]
=0.
\label{l2b}
\eeq
In the leading order in $\epsilon$ the systems
(\ref{l1a}), (\ref{l1b}) coincide with each other and yield
the first system (\ref{l2a}) while
the second one (\ref{l2b}) follows from the higher order terms.
These equations define the equilibrium locus for the
Ruijsenaars-Schneider system of particles.

{\it B) 1D Toda chain in discrete time \cite{Hirota}.}
 The reduction condition in this case is:
$$
\tau (l_1,l_2, l_3 +1, \bar l_3 +1)=
\tau (l_1,l_2, l_3 , \bar l_3 ).
$$
The discrete time 1D Toda chain in the bilinear form reads
\beq\label{1DTC}
\lambda_{13}\tau^{m-1}(x)\tau^{m+1}(x)-
\lambda_{12}\tau^{m-1}(x+\eta)\tau^{m+1}(x-\eta)
=\lambda_{23}(\tau^{m}(x))^{2}\,,
\eeq
where we have excluded $\bar l_{3}$ and have passed to the notation
of Example A). 
In this case $\vec U^{(03)}+\vec U^{(12)}$
belongs to the lattice of periods. The corresponding curve
is given by eq.\,(\ref{hyper}) with a polynomial of
{\it even} degree in the r.h.s.

The Cauchy data
$x_{i}^{0}$, $x_{i}^{1}$
with respect to the $l_3$-flow
 satisfy the  locus equations:
\beq
\prod _{s=1}^{N}\frac{
\sigma (x_{i}^{1}-x_{s}^{1}+\eta )
\sigma ^{2}(x_{i}^{1}-x_{s}^{0})}
{\sigma (x_{i}^{1}-x_{s}^{1}-\eta )
\sigma ^{2}(x_{i}^{1}-x_{s}^{0}+\eta )}
=-\, \frac{\lambda _{12}}{\lambda _{13}},
\label{l3a}
\eeq
\beq
\prod _{s=1}^{N}\frac{
\sigma (x_{i}^{0}-x_{s}^{0}+\eta )
\sigma ^{2}(x_{i}^{0}-x_{s}^{1}-\eta )}
{\sigma (x_{i}^{0}-x_{s}^{0}-\eta )
\sigma ^{2}(x_{i}^{0}-x_{s}^{1})}
=-\, \frac{\lambda _{13}}{\lambda _{12}}
\label{l3b}
\eeq

The continuous time limit can be taken similar to the way of the 
previous example. In this case, however, we have to assume
$\lambda _{13}/\lambda _{12}\rightarrow \tilde C \epsilon ^{-2}$
as $\epsilon \rightarrow 0$
with a constant $\tilde C$. We get
\beq
\sigma ^{2} (\eta )\prod _{k=1, \neq i}^{N}
\frac{\sigma (x_i -x_k +\eta )
\sigma (x_i -x_k -\eta )}
{\sigma ^{2}( x_i -x_k )}=\tilde C \dot x_{i}^{2},
\label{l4a}
\eeq
\beq
\sum _{k=1,\neq i}^{N}(\dot x_{i}+\dot x_{k})
\left [ \zeta (x_i - x_k -\eta ) +\zeta (x_i -x_k +\eta )
-2\zeta (x_i -x_k ) \right ]
=0.
\label{l4b}
\eeq
These equations follow also from
eqs.\,(4.58), (4.59) of the paper \cite{kz}.
They define the stable locus for the
RS system with respect to another
flow than in eqs.\,(\ref{l2a}), (\ref{l2b}).

{\it C) Discrete sine-Gordon equation\footnote{This version
of the discrete SG equation is different from the one
considered in \cite{HirSG}, \cite{Volkov}. The latter is closer
to a special degeneracy of the discrete KdV at 
$\lambda _{12}\rightarrow \lambda _{13}$.}.}
 The reduction condition is
$$
\tau (l_{1},l_{2},l_3 +1, \bar l_3)=
\tau (l_{1}+2,l_{2}, l_3 , \bar l_3 +1 ),
$$
so, passing to the same independent variables as in 
the previous examples, we get the equation 
\beq\label{SG}
\lambda_{13}\tau^{m}(x-\eta)\tau^{m}(x+\eta)-
\lambda_{23}\tau^{m-1}(x+\eta)\tau^{m+1}(x-\eta)
=\lambda_{12}(\tau^{m}(x))^{2}\,.
\eeq

Now it is the vector $\vec U^{(01)}+\vec U^{(32)}$
that belongs to the lattice of periods. The continuous SG equation is
reproduced in the limit $P_3 \rightarrow P_0$,
$P_{2} \rightarrow P_1$.
The spectral curves are again hyperelliptic.

 The  locus equations for
the Cauchy data
$x_{i}^{0}$, $x_{i}^{1}$
with respect to the $m$-flow are
\beq
\prod _{s=1}^{N}\frac{
\sigma (x_{i}^{0}-x_{s}^{1} )
\sigma (x_{i}^{0}-x_{s}^{1} -2\eta )}
{\sigma ^{2}(x_{i}^{0}-x_{s}^{1}-\eta )}
=\frac{\lambda _{12}}{\lambda _{13}},
\label{l5a}
\eeq
\beq
\prod _{s=1}^{N}\frac{
\sigma (x_{i}^{1}-x_{s}^{0} )
\sigma (x_{i}^{1}-x_{s}^{0}+2\eta )}
{\sigma ^{2}(x_{i}^{1}-x_{s}^{0}+\eta )}
=\frac{\lambda _{12}}{\lambda _{13}}.
\label{l5b}
\eeq
Note that the structure of these equations is different
compared to the previous examples.

\subsection{Remarks on trigonometric solutions}

Trigonometric solutions are degenerations of the elliptic
solutions when one of the periods tends to infinity. They form
a particular subfamily in the variety of
soliton solutions described in Sect.\,2.4.
The trigonometric solutions admit a very explicit description
in terms of the data defining the singular curve.

Let us set the period to be $2\pi$:
$$
\tau ^{m}(x+{2\pi })=\tau ^{m}(x)\,,
$$
so an elliptic polynomial becomes a Laurent
polynomial in $e^{ix}$. The Bethe-like equations on motion (\ref{e9a})
preserve its form, but the Weierstrass function $\sigma(x)$ is replaced by 
$\sin x$.

It follows from the periodicity that the function 
$\Psi _{0}$ (\ref{germ}) obeys
$$
\frac{\Psi _{0}(\eta x+{2\pi \eta}, m;p_j )}
{\Psi _{0}(\eta x+{2\pi \eta}, m;q_j )}
=\frac{\Psi _{0}(\eta x, m;p_j )}
{\Psi _{0}(\eta x, m;q_j )},\;\;\;\;\;\;j=1,2, \ldots , N,
$$
or, explicitly 
\beq
\frac{p_j -z_1 }{p_j - z_0 }
=\frac{q_j -z_1 }{q_j - z_0 }e^{-i\eta J_{j}},
\label{rem1}
\eeq
where $J_j$ are integer numbers. This  condition restricts admissible 
singular curves (Riemann spheres
with $N$ double points).
Minimal Laurent
polynomials correspond to the choice $J_j =\pm 1$.
This gives $N$ conditions for
$p_j , q_j$, so the number of continuous parameters in the trigonometric
case is
$2N+1$ -- the same as for the non-degenerate curves of genus $N$.

Here we do not discuss the trigonometric degeneration of the
${\cal L}$-${\cal M}$-pair and
dependence of 
of $p_j$ and $q_j$ on initial data for Ruijsenaars-Schneider
particles. This can be done along the
lines of the paper \cite{bab}.

The trigonometric loci can be  
characterized alternatively by imposing a
relation on $p_j$ and $q_j$ in addition to (\ref{rem1}):
\beq
E(p_i )=E(q_i ),\;\;\;\;\;\;\;\;
p_i \neq q_i,
\;\;\;\;\;\;\;\;i=1,2 , \ldots , N.
\label{rem2}
\eeq
For the examples A)-C) of Sect.\,3.4 the functions $E(z)$ are
\beq
\label{E(z)}
\begin{array}{l}
A)\;\;\;\;\;\;
\displaystyle{
E(z)=\frac{(z-z_2 )(z-z_3 )}{(z-z_0 )^{2}}}\,,
\\ \\
B)\;\;\;\;\;\;
\displaystyle{
E(z)=\frac{(z-z_2 )(z-z_3 )}{(z-z_1 )(z-z_0 )}}\,,
\\ \\
C)\;\;\;\;\;\;
\displaystyle{
E(z)=\frac{(z-z_1 )(z-z_2 )}{(z-z_3 )(z-z_0 )}}\,.
\end{array}
\eeq
Conditions
(\ref{rem1}), (\ref{rem2}) 
leave us with a discrete set of admissible pairs $p_i , q_i$.
The continuous parameters $\gamma _{s}$ give then an
implicit parametrization of the  loci.

\section{Conclusion}

We have shown that the main body of finite-gap theory
and the theory of elliptic solutions to nonlinear
integrable equations is also applicable to 
finite-difference (discrete) integrable equations.
 Discrete equations includes the continuous theory 
 as the result of
a limiting procedure.  Furthermore, discrete equations
 reveal some symmetries lost in the
 continuum limit. 
 We have shown that all elliptic solutions with 
 a constant number
 of zeros in the evolution (compare to \cite{KLWZ}),
  are of the algebro-geometric type. Moreover, their
 algebraic curves are spectral curves for $L$-operators of the 
Ruijsenaars-Schneider model. Each point of this curve
gives rise to 
discrete time  dynamics of zeros of the $\tau$-function.

The structure of equilibrium  loci equations of reductions of 
Hirota's equation (analogues of the
 known KdV-locus (\ref{locus})) is expected to be richer than in the
continuous case and requires further study.
It would be very interesting to extend the algebro-geometric
approach to elliptic solitons of KdV of ref.\,\cite{TV}
to the difference case as well as to  understand difference elliptic
solitons in terms of the Weierstrass reduction theory \cite{BBEIM}.

To the two main motivations pointed out at the  beginning
of the paper we can now add yet another one: an intriguing  
intimate connection
between the elliptic solutions to soliton equations
 and quantum integrable models solved by the
Bethe Ansatz. 
In our opinion, the very fact that the zeros dynamics and
equilibrium loci are described by Bethe-like equations
is remarkable and suggests  hidden parallels
between quantum and classical integrable equations.

\section*{Acknowledgments}

We thanks Ovidiu Lipan for discussions and for his interest 
in the subject and J.Talstra for help.
The work of I.K. was supported by RFBR grant 95-01-00751. 
The work of A.Z. was supported in part by RFBR grant
97-02-19085, by ISTC grant 015 and also by NSF grant DMR-9509533.
A.Z. is grateful to Professor R.Seiler for the kind hospitality
at the Technische Universit\"at Berlin where a part of this
work was done. P.W was supported by NSF grant DMR-9509533.
P.W and A.Z thank Institute for Theoretical Physics at Santa Barbara for
its hospitality in April 1997 where this paper has been completed.

\end{document}